\documentclass[pra,aps,superscriptaddress,nofootinbib,twocolumn]{revtex4-1}
\pdfoutput=1
\usepackage{mathrsfs}
\usepackage{amsfonts}
\usepackage{amssymb}
\usepackage{amsmath}
\usepackage{graphicx}
\usepackage[usenames,dvipsnames]{color}
\usepackage{ulem}
\usepackage{lmodern}
\usepackage{amsthm}
\usepackage{xcolor}
\usepackage{subfigure}
\usepackage{url}
\usepackage{hyperref}

\hypersetup{
    colorlinks=true,
    linkcolor=blue,
    filecolor=blue,      
    urlcolor=blue,
    citecolor=cyan,
}

\newcommand{\Tr}{\mathrm{Tr}}
\newcommand{\norm}[1]{\Vert #1 \Vert}

\begin{document}

\title{Perturbative tomography of small errors in quantum gates}

\author{Ruyu Yang}
\affiliation{Graduate School of China Academy of Engineering Physics, Beijing 100193, China}

\author{Ying Li}
\email{yli@gscaep.ac.cn}
\affiliation{Graduate School of China Academy of Engineering Physics, Beijing 100193, China}

\begin{abstract}
We propose an efficient protocol to fully reconstruct a set of high-fidelity quantum gates. Usually, the efficiency of reconstructing high-fidelity quantum gates is limited by the sampling noise. Our protocol is based on a perturbative approach and has two stages. In the first stage, the unital part of noisy quantum gates is reconstructed by measuring traces of maps, and the trace can be measured by amplifying the noise in a way similar to randomised benchmarking and quantum spectral tomography. In the second stage, by amplifying the non-unital part using the unital part, we can efficiently reconstruct the non-unital part. We show that the number of measurements needed in our protocol scales logarithmically with the error rate of gates. 
\end{abstract}

\maketitle

\section{Introduction}

Quantum computing can solve many problems that are intractable for classical computing. In the standard circuit model of universal quantum computing, all quantum algorithms can be realised by combining elementary unitary evolutions, i.e.~quantum gates~\cite{nielsen2002quantum}. In the past twenty years, the fidelity of quantum gates have been constantly improved. In superconducting and trapped-ion systems, single-qubit gate fidelities have achieved 99.9\%~\cite{barends2014superconducting} and 99.9999\%~\cite{harty2014high}, respectively. However, these fidelities are not sufficiently low for directly implementing large-scale quantum algorithms, e.g.~Shor's algorithm~\cite{beauregard2002circuit}. Methods such as the quantum error correction~\cite{steane1996error,calderbank1997quantum,fowler2012surface, o2017quantum,chiaverini2004realization,reed2012realization} and mitigation protocols~\cite{li2017efficient, temme2017error, endo2018practical} have been proposed and demonstrated~\cite{kandala2019error,song2019quantum,zhang2020error} to minimise the impact of errors in the quantum computing. Quantum gate characterisation is of importance to debug the gates and take the full advantage of these error correction methods. The common approaches of gate characterisation include measuring the average fidelity through randomized benchmarking~\cite{emerson2005scalable,knill2008randomized, magesan2011scalable, magesan2012characterizing, sheldon2016characterizing, fong2017randomized, proctor2017randomized, wallman2018randomized, onorati2019randomized,lu2015experimental,mckay2019three} and reconstructing all the detailed information using quantum tomography~\cite{chuang1997prescription,poyatos1997complete,d2001quantum,altepeter2003ancilla,mohseni2006direct,merkel2013self,blume2013robust,stark2014self,greenbaum2015introduction,blume2017demonstration,sugiyama2018reliable}. Quantum spectral tomography is recently proposed to obtain eigenvalues of quantum gates~\cite{helsen2019spectral}. 

In this paper, we describe a method of characterising noisy gates that are closed to perfect unitary gates. The unital part of the completely positive map describing a noisy gate can be reconstructed by measuring the traces of maps, e.g.~using randomised benchmarking~\cite{kimmel2014robust}. We propose a way to measure the trace using deterministic gate sequences inspired by the quantum spectral tomography~\cite{helsen2019spectral}. In the trace measurement, the fidelity decreases slower with the sequence length in deterministic gate sequences compared with random sequences. Therefore, we can use sufficiently long sequences to amplify the error for the efficient measurement~\cite{blume2017demonstration}. The reconstruction of the unital part is based on a perturbative approach, we express the error in a gate as a perturbation and neglecting high-order effects of the error in the data analysis. With the unital part reconstructed, we can amplify and efficiently measure the non-unital part in a similar way. 

The obstacle of high-fidelity-gate reconstruction is the sampling noise. In order to reconstruct a noisy gate, we need to suppress the sampling noise to a level that is lower than the error rate. Our method inherits the advantage of randomized benchmarking and quantum spectral tomography, that the error is amplified in a long gate sequence to reduce the sampling noise~\cite{blume2017demonstration}. In our protocol, the problem of state preparation and measurement errors is overcome as the same as in the quantum gate set tomography (GST)~\cite{merkel2013self,blume2013robust,stark2014self,greenbaum2015introduction,blume2017demonstration,sugiyama2018reliable}. We focus on the case of one-qubit gates in this paper, and the method can be generalised to multi-qubit gates. The correlated errors in multi-qubit systems can be characterised using the perturbative tomography protocol proposed recently~\cite{govia2020bootstrapping}. We find that the number of measurements needed for sufficiently low sampling noise scales logarithmically with the error rate of gates. Therefore, this work paves an efficient way for the quantum tomography of high-fidelity gates. 

This paper is organized as follows. In Sec.~\ref{sec:maps} we give a brief review on completely positive maps, including the Pauli transfer matrix representation of maps~\cite{greenbaum2015introduction}. In Sec.~\ref{sec:accumulation}, we discuss the error accumulation in a deterministic gate sequence. In Sec.~\ref{sec:protocol}, we present the method for trace measurement, and reconstructions of unital and non-unital parts. In Sec.~\ref{sec:num}, we numerically demonstrate our protocol with the finite sampling noise. Conclusions are given in In Sec.~\ref{sec:conclusion}. 

\section{Quantum maps}
\label{sec:maps}

The completely positive map describes the evolution of a quantum system without initial correlation between the system and environment~\cite{choi1975completely, jordan2004dynamics}, which can be written in the operator-summation form~\cite{nielsen2002quantum}: 
\begin{eqnarray}
\mathcal{M}(\rho) = \sum_q K_q \rho K_q^\dag.
\end{eqnarray}
The Kraus operators satisfy $\sum_q K_q^\dag K_q = \openone$ if the map is trace-preserving, where $\openone$ is the identity operator. The Pauli transfer matrix of a map reads 
\begin{eqnarray}
M_{\sigma,\tau} = d^{-1}\Tr\left[\sigma\mathcal{M}(\tau)\right],
\end{eqnarray}
which is the matrix representation of the map using Pauli operators as the basis of the operator space, according to the Hilbert-Schmidt inner product. Here, $\sigma$ and $\tau$ are Pauli operators, and $d$ is the dimension of the Hilbert space. We can find that all elements of the Pauli transfer matrix are real, and $M_{\sigma,\tau} \in [-1,1]$. Let $M_j$ be the Pauli transfer matrix of the map $\mathcal{M}_j$, then the matrix of $\mathcal{M}_i \mathcal{M}_j$ is $M_i M_j$. 

We always take the identity operator as the first element in Pauli operators. The first row of the matrix is $M_{\openone,\sigma} = \delta_{\openone,\sigma}$ for a trace-preserving map. Therefore, we can write the matrix of a trace-preserving map in the form 
\begin{eqnarray}
M = \left[
\begin{array}{cc}
1 & \vec{0}^{\rm \, T}\\
\vec{k} & E
\end{array}
\right]
\label{eq:M}
\end{eqnarray}
The matrix is $d^2$-dimensional in general. In this paper, we only consider the case of one qubit, i.e.~$d = 2$. Then, $\vec{0}$ and $\vec{k}$ are three-dimensional column vectors, all elements of $\vec{0}$ are zero, and $E$ is a three-dimensional matrix. If the map is unital, $\vec{k} = \vec{0}$. We call $E$ the unital part and $\vec{k}$ the non-unital part. When the map is completely positive, there is a constraint on $E$ and $\vec{k}$, which is~\cite{rudnicki2018gauge} 
\begin{equation}
\Vert \vec{k} \Vert^2 \leq 1-\vert \lambda_1 \vert^2-\vert \lambda_2 \vert^2-\vert \lambda_3 \vert^2+2\lambda_1 \lambda_2 \lambda_3,
\end{equation}
where $\lambda_l$ are eigenvalues of $E$. 

An ideal quantum gate is a unitary evolution in the form $\mathcal{M}^{\rm i}(\rho) = U\rho U^\dag$, where $U$ is the unitary operator. We use $M^{\rm i}$ to denote the Pauli transfer matrix of the ideal gate $\mathcal{M}^{\rm i}$, which is always a unitary matrix. Let $E^{\rm i}$ and $\vec{k}^{\rm i}$ be the unital and non-unital parts of $M^{\rm i}$, then $E^{\rm i}$ is a unitary matrix, all eigenvalues of $E^{\rm i}$, i.e.~$\lambda^{\rm i}_l$, have the same absolute value of $1$, and $\vec{k}^{\rm i} = \vec{0}$. Accordingly, for a quantum gate with high-fidelity, absolute values $\vert\lambda_l\vert$ are all close to $1$, and the non-unital part $\vec{k}$ is close to zero. 

The error in a quantum gate is the difference between the actual noisy gate and the ideal gate, i.e.
\begin{eqnarray}
\delta M = M - M^{\rm i}.
\end{eqnarray}
When the gate fidelity is high, $\delta M$ must be close to zero. 

In this paper, we will consider a set of quantum gates. We use the subscript to label the gate, i.e.~$M_j$, $M_j^{\rm i}$ and $\delta M_j$ are respectively the actual noisy matrix, ideal matrix and error of the gate-$j$. The error can be gate dependent. We assume that the error is time-independent and uncorrelated~\cite{huo2018self}. 

\section{Error accumulation}
\label{sec:accumulation}

To efficiently measure the small error in a quantum gate, we can repeat the noisy gate such that the error accumulates with the repetition length. If the gate $\mathcal{M}$ is repeated for $n$ times, the corresponding Pauli transfer matrix reads 
\begin{eqnarray}
M^{n} = \left[
\begin{array}{cc}
1 & \vec{0}^{\rm \, T}\\
(\sum_{q=0}^{n-1} E^q)\vec{k} & E^{n}
\end{array}
\right].
\label{eq:Mk}
\end{eqnarray}

The magnitude of the unital part decreases exponentially with the repetition length $n$, i.e.~$E^n\sim \lambda_l^n$. If the gate is of high-fidelity, the eigenvalue $\vert \lambda_l \vert = 1-\epsilon$ is close to $1$. Then, we can take $n\sim 1/\epsilon$ such that $E^n$ is significantly changed by the accumulated error. If $E^n$ is measured with the accuracy $\eta$, we can estimate the error in the unital part of one gate with the accuracy $\sim \eta/n$, i.e.~$\sim \eta\epsilon$. 

By repeating the gate, the non-unital part is amplified by $\sum_{q=0}^{n-1} E^q$. In three eigenvalues of $E$, one of them (the eigenvalue itself rather than the absolute value) is always close to $1$, if the fidelity is high. Without loss of generality, we assume that $\lambda_1 = 1-\epsilon'$. Then, $\sum_{q=0}^{n-1} E^q \sim 1/\epsilon'$ in the limit $n\rightarrow\infty$. If the non-unital part of $M^n$ is measured with the accuracy $\eta'$, we can estimate the error in the non-unital part of one gate with the accuracy $\sim \eta'\epsilon'$. 

Later, we will show how to reconstruct a noisy gate efficiently by accumulating the error. We will find that in the repeated gate, the trace of map is the robust information that can be extracted with high accuracy, and the estimation of individual eigenvalues is not robust. In the randomised benchmarking, the trace of the product of the noisy gate and an ideal Clifford gate can be measured, in which the trace is related to the relative fidelity between the noisy gate and the ideal Clifford gate~\cite{kimmel2014robust}. The repetition length is limited by the relative fidelity in the randomised benchmarking. To reconstruct the noisy gate, we need to measure traces of a set of products, and it is impossible that relative fidelities are high for all of them. In our case, the repetition length is limited by the eigenvalues. As long as the gate is close to a unitary gate, the absolute values of eigenvalues are close to $1$, and a large repetition length is permitted. 

\section{Protocol}
\label{sec:protocol}

The protocol has two stages. In the first stage, the unital part of gates is reconstructed by using trace measurements. In the second stage, the non-unital part is reconstructed by amplifying them using the unital part. We present our protocol as follows. 

\subsection{Trace measurement}
\label{sec:trace}

\begin{figure}[tbp]
\centering
\includegraphics[width=1\linewidth]{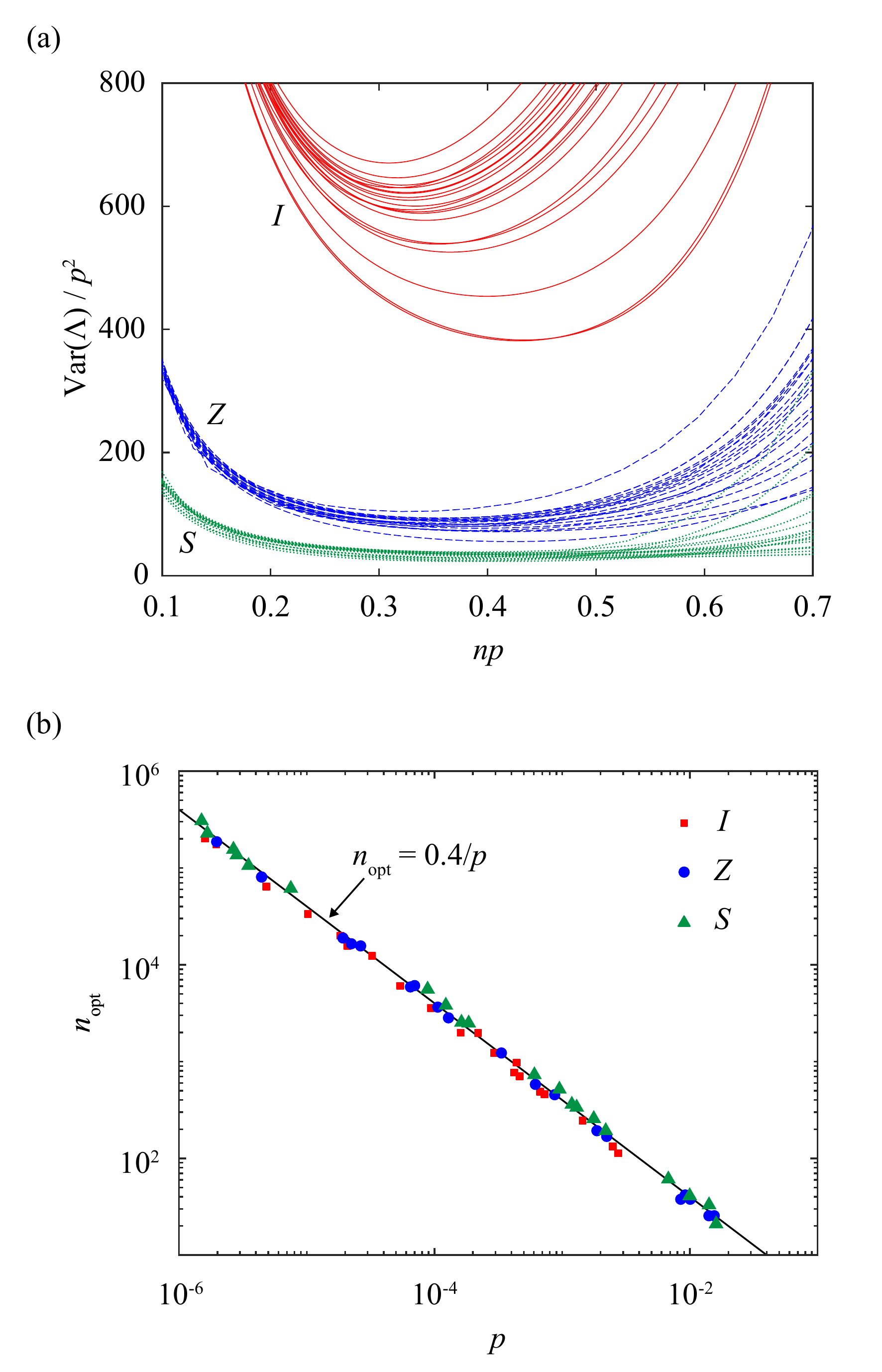}
\caption{
(a) The variance as a function of $n$. (b) The optimal value of $n$ (i.e.~$n_{\rm opt}$) that minimizes the variance in (a). Each curve in (a) corresponds to a gate with randomly generated error (see Appendix~\ref{app:error}). The error rate $p$ is randomly selected in the range $\sim 10^{-2} - 10^{-6}$. 
}
\label{fig:trace_variance}
\end{figure}

In the experiment, it is difficult to observe the effect of low-level noise such as in high-fidelity gates. To characterise the noise, we can repeat the noisy gate to accumulate the errors, similar to the randomised benchmarking~\cite{knill2008randomized,magesan2011scalable,magesan2012characterizing,sheldon2016characterizing,fong2017randomized,onorati2019randomized,wallman2018randomized,proctor2017randomized} and quantum spectral tomography~\cite{helsen2019spectral}. The matrix of the repeated gate is $M^n$, see Eq.~(\ref{eq:Mk}). The eigenvalues of $M^n$ are $1$, $\lambda_1^n$ ,$\lambda_2^n$ and $\lambda_3^n$, where $\lambda_1$, $\lambda_2$ and $\lambda_3$ are eigenvalues of $E$. We note that $M$ and $M^n$ can always be expressed in the Jordan normal form, which leads to the formula 
\begin{equation}
\Tr(M^n) = 1 + \lambda_1^n + \lambda_2^n + \lambda_3^n
\end{equation}

The protocol for measuring the trace is as follows: 
\begin{itemize}
\item[$\bullet$] Use GST to obtain an estimate of $M^{l}$, and the estimate is $\hat{M}_{l}$, where $l = n,2n,3n$; compute the trace of the unital part $t_{l} = \Tr(\hat{M}_{l})-1$ for each $l$; 
\item[$\bullet$] Solve the system of equations(SOE) to obtain $\lambda_1$, $\lambda_2$ and $\lambda_3$; compute $\Lambda = 1+\lambda_1+\lambda_2+\lambda_3$, which is the estimate of the trace of the map $\Tr(M)$.
\begin{subequations}
\begin{align}
\lambda_1^{n} + \lambda_2^{n} + \lambda_3^{n} &= t_{n}, \\
\lambda_1^{2n} + \lambda_2^{2n} + \lambda_3^{2n} &= t_{2n}, \\
\lambda_1^{3n} + \lambda_2^{3n} + \lambda_3^{3n} &= t_{3n}
\end{align}
\label{eq:trace}
\end{subequations}

\end{itemize}

In GST, because of the state preparation and measurement errors, the estimate and the actual matrix are related by an unknown similarity transformation, i.e.~$M^{l} = T \hat{M}_{l} T^{-1}$, assuming the sampling error in GST is negligible~\cite{merkel2013self, greenbaum2015introduction, blume2017demonstration, rudnicki2018gauge,stark2014self,sugiyama2018reliable,blume2013robust}. Although the transformation is unknown, the trace can be directly obtained using the estimate, i.e.~$\Tr(M^{l}) = \Tr(T \hat{M}_{l} T^{-1}) = \Tr(\hat{M}_{l})$, i.e.~the trace measurement is robust to the state preparation and measurement errors. 

In our protocol, the eigenvalues are computed by solving SOE. Note that there are multiple solutions of SOE~(\ref{eq:trace}). These solutions are close to each other when $n$ is large, and the difference between them is typically $O(2\pi/n)$. However, only one of them is correct. In order to efficiently identify the correct solution, we can implement the trace measurement (to compute eigenvalues) for a monotonically increasing series of $n$, i.e.~$n=n_1,n_2,n_3,\ldots,n_{\rm max}$. For each value of $n$, we construct and solve SOE as in the protocol. We always take $n_1 = 1$ such that solutions of $n = n_1$ are significantly different. Then, under the assumption that the error is small, we can rule out solutions that are far from eigenvalues of $M^{\rm i}$. For $n=n_{i>1}$, we choose the solution that is the closest to the solution of $n=n_{i-1}$. In this way, we only need to reduce the sampling noise at $n=n_{i-1}$ to the level that the confidence interval is sufficiently small for distinguishing solutions of $n = n_i$. In the numerical demonstrations, we will show that such a procedure is efficient by taking a power series of $n$. 

\begin{table}
\begin{tabular}{|c|c|}
\hline
$U_1 = e^{\frac{i\pi}{6}\hat{Z}}$ & $U_2 = e^{{\frac{-i\pi}{3}}\frac{\hat{X}+\hat{Y}-\hat{Z}}{\sqrt{3}}}$ \\
\hline
$U_3 = e^{{\frac{-2i\pi}{3}}\frac{\hat{X}+\hat{Y}-\hat{Z}}{\sqrt{3}}}$ & $U_4 = e^{{\frac{-i\pi}{3}}\frac{\hat{X}+\hat{Y}+\hat{Z}}{\sqrt{3}}}$ \\
\hline
$U_5 = e^{{\frac{-2i\pi}{3}}\frac{\hat{X}+\hat{Y}+\hat{Z}}{\sqrt{3}}}$ & $U_6 = e^{{\frac{-i\pi}{3}}\frac{\hat{X}-\hat{Y}+\hat{Z}}{\sqrt{3}}}$ \\
\hline
$U_7 = e^{{\frac{-2i\pi}{3}}\frac{\hat{X}-\hat{Y}+\hat{Z}}{\sqrt{3}}}$ & \\
\hline
\end{tabular}
\caption{
The gate set used in the numerical demonstration. The ideal map of the gate $U_i$ is $\mathcal{M}_i^{\rm i}(\rho) = U_i\rho U_i^\dag$. 
}
\label{table:gate_set}
\end{table}

Finally, the trace is computed using eigenvalues obtained at $n=n_{\rm max}$. We need to choose a sufficiently large $n_{\rm max}$ in order to amplify the noise. Later we show that the optimal value of $n_{\rm max}$ is $\sim 0.4/p$. Here, $p = 1-F$ is the error rate, and $F$ is the average fidelity~\cite{nielsen2002quantum}. We remark that we can also use the method of least squares~\cite{wolberg2006data} and the matrix pencil method~\cite{sarkar1995using} to work out the eigenvalues and then compute the trace. 
\subsubsection*{Variance of the trace measurement}

The variance of $\Lambda(t_{n},t_{2n},t_{3n})$ is 
\begin{eqnarray}
{\rm Var}(\Lambda) \approx \sum_{l=n,2n,3n} \left(\frac{\partial\Lambda}{\partial t_{l}}\right)^2 {\rm Var}[t_{l}]
\end{eqnarray}
where 
\begin{widetext}
\begin{subequations}
\begin{align}
\frac{\partial\Lambda}{\partial t_{n}} &= \frac{\lambda_1^{-n}\lambda_2^{-n}\lambda_3^{-n}(\lambda_1\lambda_2^{2n}\lambda_3^{2n}(\lambda_2^n-\lambda_3^n)+\lambda_1^{3n}(\lambda_2^{2n}\lambda_3-\lambda_2\lambda_3^{2n})+\lambda_1^{2n}(-\lambda_2^{3n}\lambda_3+\lambda_2\lambda_3^{3n}))}{(\lambda_1^n-\lambda_2^n)(\lambda_1^n-\lambda_3^n)(\lambda_2^n-\lambda_3^n)n}  \\
\frac{\partial\Lambda}{\partial t_{2n}} &= \frac{\lambda_1^{-n}\lambda_2^{-n}\lambda_3^{-n}(\lambda_1^{3n}(-\lambda_2^n\lambda_3+\lambda_2\lambda_3^n)+\lambda_1^{n}(\lambda_2^{3n}\lambda_3-\lambda_2\lambda_3^{3n})+\lambda_1^{}(-\lambda_2^{3n}\lambda_3^n+\lambda_2^n\lambda_3^{3n}))}{2(\lambda_1^n-\lambda_2^n)(\lambda_1^n-\lambda_3^n)(\lambda_2^n-\lambda_3^n)n} \\
\frac{\partial\Lambda}{\partial t_{3n}} &= \frac{\lambda_1^{-n}\lambda_2^{-n}\lambda_3^{-n}(\lambda_1\lambda_2^{n}\lambda_3^{n}(\lambda_2^n-\lambda_3^n)+\lambda_1^{2n}(\lambda_2^{n}\lambda_3-\lambda_2\lambda_3^{n})+\lambda_1^{n}(-\lambda_2^{2n}\lambda_3+\lambda_2\lambda_3^{2n}))}{3(\lambda_1^n-\lambda_2^n)(\lambda_1^n-\lambda_3^n)(\lambda_2^n-\lambda_3^n)n}
\end{align}
\end{subequations}

When $\lambda_1 \neq \lambda_2 = \lambda_3$, we have 
\begin{subequations}
\begin{align}
\lim_{\lambda_2 \to \lambda_3}\frac{\partial \Lambda}{\partial t_{n}} &= \frac{\lambda_1^{-n}\lambda_3^{-n}(\lambda_1^{2n}\lambda_3^{1+n}(1-3n)+\lambda_1\lambda_3^{3n}n+\lambda_1^{3n}\lambda_3(-1+2n))}{(\lambda_1^n-\lambda_3^n)^2n^2} \\
\lim_{\lambda_2 \to \lambda_3}\frac{\partial \Lambda}{\partial t_{2n}} &= \frac{\lambda_1^{-n}\lambda_3^{-2n}(-\lambda_1^{3n}\lambda_3(-1+n)+\lambda_1^{n}\lambda_3^{1+2n}(-1+3n)-2\lambda_1\lambda_3^{3n}n)}{2(\lambda_1^n-\lambda_3^n)^2n^2} \\
\lim_{\lambda_2 \to \lambda_3}\frac{\partial \Lambda}{\partial t_{3n}} &= \frac{\lambda_1^{-n}\lambda_3^{-2n}(\lambda_1^{n}\lambda_3^{1+n}(1-2n)+\lambda_1^{2n}\lambda_3(-1+n)+\lambda_1\lambda_3^{2n}n)}{3(\lambda_1^n-\lambda_3^n)^2n^2}
\end{align}
\end{subequations}

When $\lambda_1 = \lambda_2 = \lambda_3$, we have 
\begin{subequations}
\begin{align}
\lim_{\lambda_2,\lambda_3 \to \lambda_1}\frac{\partial\Lambda}{\partial t_{n}} &= \frac{\lambda_1^{1-n}(1-5n+6n^2)}{2n^3} \\
\lim_{\lambda_2,\lambda_3 \to \lambda_1}\frac{\partial\Lambda}{\partial t_{2n}} &= \frac{\lambda_1^{1-2n}(1-4n+3n^2)}{2n^3} \\
\lim_{\lambda_2,\lambda_3 \to \lambda_1}\frac{\partial\Lambda}{\partial t_{3n}} &= \frac{\lambda_1^{1-3n}(-1+n)(-1+2n)}{6n^3}
\end{align}
\end{subequations}
\end{widetext}

Therefore, the variance is always convergent even if eigenvalues are degenerate. We remark that in the case $\lambda_1 \neq \lambda_2 = \lambda_3$, we need to avoid the value of $n$ with $\lambda_1^n \approx \lambda_3^n$. 

The variance is plotted in Fig.~\ref{fig:trace_variance} for quantum gates with randomly generated errors. The identity gate $I$, Pauli gate $Z$ and phase gate $S$ are considered, corresponding to the cases $\lambda_1 = \lambda_2 = \lambda_3$, $\lambda_1 \neq \lambda_2 = \lambda_3$ and $\lambda_1 \neq \lambda_2 \neq \lambda_3$, respectively. For each ideal gate, twenty noisy gates are generated by computing the time integral of randomly generated Lindblad superoperator (see Appendix~\ref{app:error}). We take $n = 4l+1$, where $l$ is an integer, such that ${\lambda_1}^n \neq {\lambda_3}^n$ for the gate $S$. We can find that the variance is minimised around $n \sim 0.4/p$. The gate $I$ has the highest variance, and the gate $S$ has the lowest variance in the three gates. 

\subsection{Unital part reconstruction}

In our protocol, we reconstruct the unital part by measuring the trace of actual noisy gates using the method given in Sec.~\ref{sec:trace}. According to Ref.~\cite{kimmel2014robust}, we can also reconstruct the unital part of the map $M$ by measuring the trace $\Tr(C^{\rm i}M)$, where $C^{\rm i}$ is one of ideal Clifford gates, and the trace can be measured using the randomised benchmarking. 

We use the perturbative approach. For the gate-$j$, we express the Pauli transfer matrix of the actual noisy gate as $M_j = M_j^{\rm i} + \delta M_j$. Because our aim is to reconstruct the map rather than measuring the relative fidelity with respect to an ideal gate, $M_j^{\rm i}$ is up to choice, however, must be close to the actual noisy gate $M_j$. The error can be written as 
\begin{eqnarray}
\delta M_j = \left[
\begin{array}{cc}
0 & \vec{0}^{\rm \, T}\\
\delta \vec{k}_j & \delta E_j
\end{array}
\right].
\end{eqnarray}
In this section, we show how to reconstruct $\delta E_j$. 

For a quadruple map $\mathcal{M}_{i,j,k,l} = \mathcal{M}_{i}\mathcal{M}_{j}\mathcal{M}_{k}\mathcal{M}_l$, the Pauli transfer matrix is 
\begin{eqnarray}
&& M_{i,j,k,l} = M_i M_j M_k M_l \notag \\
&=& M_i^{\rm i} M_j^{\rm i} M_k^{\rm i} M_l^{\rm i} + \delta M_i M_j^{\rm i} M_k^{\rm i} M_l^{\rm i} + M_i^{\rm i} \delta M_j M_k^{\rm i} M_l^{\rm i} \notag \\
&&+ M_i^{\rm i} M_j^{\rm i} \delta M_k M_l^{\rm i} + M_i^{\rm i} M_j^{\rm i} M_k^{\rm i} \delta M_l + O(\delta^2).
\end{eqnarray}
By measuring the trace of quadruple maps, we are able to obtain the unital part of each $\delta M_j$. We remark that $\Tr(M_i^{\rm i} M_j^{\rm i} M_k^{\rm i} \delta M_l) = \Tr(E_i^{\rm i} E_j^{\rm i} E_k^{\rm i} \delta E_l )$. 

The protocol for reconstructing the unital part is as follows: 
\begin{itemize}
\item[$\bullet$] Given a set of gates $\{M_j\}$, measure the trace of quadruple maps $\Tr(M_{i,j,k,l})$ using our protocol; 
\item[$\bullet$] Solve the SOE for a set of quadruple maps, 
\begin{eqnarray}
\Tr(M_{i,j,k,l}) &=& \Tr(M_i^{\rm i} M_j^{\rm i} M_k^{\rm i} M_l^{\rm i}) + \Tr(E_j^{\rm i} E_k^{\rm i} E_l^{\rm i} \delta E_i) \notag \\
&&+ \Tr(E_k^{\rm i} E_l^{\rm i} E_i^{\rm i} \delta E_j) + \Tr(E_l^{\rm i} E_i^{\rm i} E_j^{\rm i} \delta E_k) \notag \\
&&+ \Tr(E_i^{\rm i} E_j^{\rm i} E_k^{\rm i} \delta E_l)
\label{eq:quadruple}
\end{eqnarray}
to obtain each element of $\delta E_j$; 
\item[$\bullet$] Iterate the second step by replacing $M_j^{\rm i}$ with $M_j^{\rm i} + \delta E_j$. 
\end{itemize}
The iteration can rapidly increase the accuracy by taking into account higher-order effects, which is not necessary when $\delta E_j$ is sufficiently small. In our numerical simulation that we will show later, the iteration is not used. 

\begin{table}
\begin{tabular}{|c|c|c|c|c|c|c|c|}
\hline
5,2,3,6 & 7,2,2,6 & 5,1,4,3 & 6,3,4,1 & 6,1,4,6 & 1,5,3,5 & 6,1,2,5 & 5,5,2,6 \\
\hline
4,2,1,4 & 3,2,3,4 & 4,7,2,7 & 2,1,2,3 & 1,7,1,2 & 7,3,6,3 & 7,4,1,7 & 5,6,4,6 \\
\hline
3,1,3,3 & 5,1,5,1 & 1,2,4,3 & 3,6,1,4 & 4,5,4,4 & 5,4,5,1 & 5,3,5,5 & 5,7,1,4 \\
\hline
4,1,4,6 & 2,2,2,2 & 7,1,2,3 & 1,4,7,6 & 1,1,5,1 & 6,3,6,1 & 5,1,3,2 & 6,4,4,3 \\
\hline
3,2,6,6 & 2,7,4,1 & 7,6,5,2 & 6,4,1,6 & 6,7,1,6 & 5,4,2,5 & 1,6,4,1 & 1,5,6,7 \\
\hline
2,5,2,5 & 1,1,5,6 & 7,2,6,5 & 5,6,5,7 & 1,7,6,5 & 5,2,3,2 & 2,7,6,1 & 7,2,1,5 \\
\hline
1,5,1,6 & 1,6,7,4 & 7,7,2,3 & 6,5,4,5 & 4,5,2,5 & 5,4,7,4 & 1,2,2,1 & 6,7,1,3 \\
\hline
7,4,7,5 & 7,4,2,7 & 5,4,1,2 & 2,3,3,3 & 1,1,3,7 & 1,7,6,2 & 7,6,4,4 & 5,7,4,4 \\
\hline
4,6,3,1 & 1,2,5,2 & 6,7,1,2 & 1,4,7,7 & 5,1,7,5 & 1,7,5,7 & 2,7,1,5 & 5,1,1,5 \\
\hline
2,5,4,6 & 7,4,6,3 & 1,5,6,1 & 2,3,6,5 & 5,2,5,1 & 7,5,3,3 & 5,3,7,3 & 4,1,1,7 \\
\hline
4,2,1,7 & 7,4,3,3 & 5,4,1,3 & 2,7,1,2 & 1,2,3,6 & 5,4,1,2 & 1,3,7,1 & 6,6,2,3 \\
\hline
5,1,4,2 & 1,2,2,6 & 4,4,4,5 & 1,1,6,6 & 1,7,7,6 & 4,1,2,5 & 2,2,2,1 & 4,2,7,3 \\
\hline
4,7,5,7 & 4,7,3,3 & 2,5,3,6 & 4,1,4,4 & & & & \\
\hline
\end{tabular}
\caption{
Quadruple maps for reconstructing the unital part. Here $(i,j,k,l)$ denotes the quadruple map $M_{i,j,k,l} = M_i M_j M_k M_l$. 
}
\label{table:quadruple}
\end{table}

We do not need to measure all quadruple maps. Each matrix $\delta E_n$ has nine elements. For a set of $N$ gates, the total number of matrix elements is $9N$. However, we can never find $9N$ linearly independent equations, because of the gauge problem of GST~\cite{greenbaum2015introduction,rudnicki2018gauge} i.e.~the Pauli transfer matrix can only be reconstructed up to a similarity transformation. Therefore, the maximum number of linearly independent equations is $9N-8$, where $8$ is due to the similarity transformation of three-dimensional matrices. See Appendix~\ref{app:gauge}. Therefore, we need to identify and measure at least $9N-8$ quadruple maps that provide $9N-8$ linearly independent equations. 

In Table~\ref{table:gate_set}, we list seven gates, whose quadruple maps lead to $9N-8$ linearly independent equations. A hundred quadruple maps are given in Table~\ref{table:quadruple}, and $9\times 7 - 8 = 55$ of them are linearly independent. We choose these quadruple maps because their unital parts have three different eigenvalues, in order to minimise the variance. In principle, we can also use the product of two and three maps rather than four to construct linear equations. However, we find numerically that they are insufficient for constructing $9N-8$ linearly independent equations if we only choose the double or triple maps with three different eigenvalues. This gate set is complete, and any unital map can be expressed as a linear combination of maps of these gates and their products. 

Once we have a complete set of gates reconstructed, the unital part of any other map $M'$ can be reconstructed by measuring $\Tr(M_jM')$~\cite{kimmel2014robust}. The protocol is as follows: Given a gate $M'$ and a set of $9$ linearly independent maps $\{M_j\}$ (maps of gates in the gate set and their products), measure the trace $\Tr(M_j M')$; then solve SOE 
\begin{eqnarray}
\Tr(M_j M') &=& \Tr(E_j E') + 1
\end{eqnarray}
to obtain each element of $E'$, where $E'$ is the unital part of $M'$. 

\subsection{Non-unital part reconstruction}

\begin{figure}[tbp]
\centering
\includegraphics[width=1\linewidth]{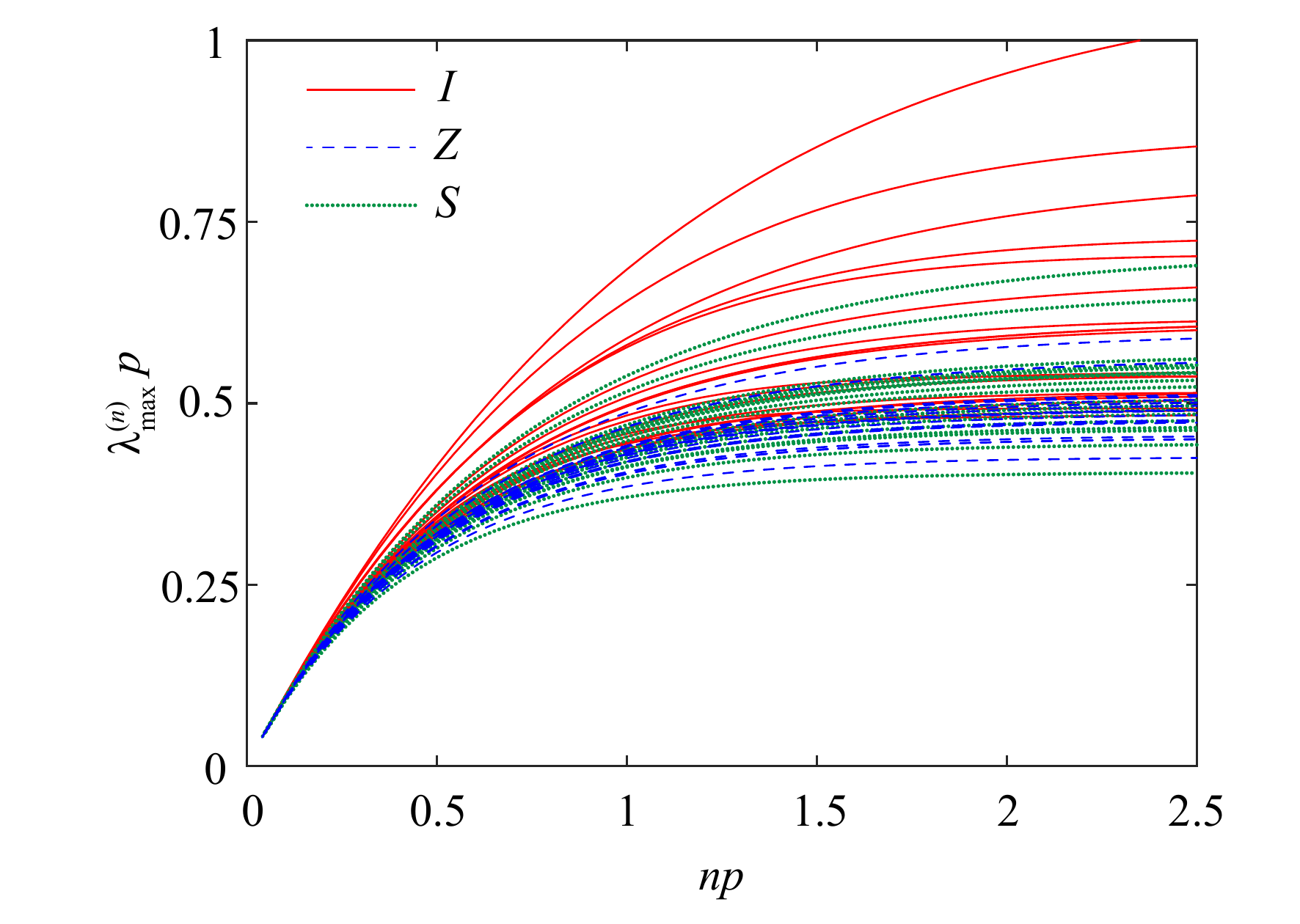}
\caption{
The largest singular value $\lambda_{\rm max}^{(n)}$ of $E^{(n)}$ as a function of number of repetitions $n$. $p$ is the error rate. Each curve corresponds to a gate with randomly generated error (see Appendix~\ref{app:error}). The error rate $p$ is randomly selected in the range $\sim 10^{-2} - 10^{-6}$. 
}
\label{fig:singular_value}
\end{figure}

Given the unital part reconstructed, we can amplify and reconstruct the non-unital part in a similar way. Repeating the map $M$ for $n$ times, the non-unital part of $M^n$ is $\vec{k}^{(n)} = E^{(n)}\vec{k}$, where $E^{(n)} = \sum_{q=0}^{n-1} E^q$ [see Eq.~(\ref{eq:Mk})]. Using the the conventional quantum tomography protocol, e.g.~GST, we can obtain the non-unital part of $M^n$ in the experiment. By solving the equation, we can compute the non-unital part of $M$, i.e.~$\vec{k} = {E^{(n)}}^{-1}\vec{k}^{(n)}$. We remark that the unital part $E$ has been reconstructed. Because $\vec{k}^{(n)}$ is directly measured in the experiment, it has a finite variance due to the sampling noise. Therefore, the variance of $\vec{k}$ depends on singular values of $E^{(n)}$. 

When the gate error is small, the unital part $E$ is close to a unitary matrix, and at least one of its eigenvalues is close to one. Without loss of generality, we suppose $\lambda_1$ is the eigenvalue close to one. Then $1 - \lambda_1 \sim p$, where $p$ is the error rate. The largest singular value of $E^{(n)}$ is $\lambda_{\rm max}^{(n)} \sim \frac{1}{1-\lambda_{1}}$, when $n$ is sufficiently large. In Fig.~\ref{fig:singular_value}, we plot the largest singular value $\lambda_{\rm max}^{(n)}$ of $E^{(n)}$ for quantum gates with randomly generated errors. We can find that $\lambda_{\rm max}^{(n)}$ approaches $\sim 1/p$ when the repetition number $n$ is sufficiently large. For other two eigenvalues, if they are not close to one, they cannot efficiently amplify the non-unital part, i.e.~reduce the variance of $\vec{k}$. Therefore, we can only make sure one component of $\vec{k}$ measured with low variance: Given $\vec{k}^{(n)}$ measured with the variance $\sigma^2_n$ and the largest singular value $\lambda_{\rm max}^{(n)}\sim 1/p$, the variance of the corresponding component is $\sim p^2\sigma^2_n$. To reconstruct all components, we need to combine maps as in the unital part reconstruction. 

The protocol for reconstructing the non-unital part is as follows: 
\begin{itemize}
\item[$\bullet$] Given a set of gates $\{M_j\}$, measure the non-unital part of repeated double maps $(M_i M_j)^n$ using GST, which is denoted by $\vec{k}_{i,j}^{(n)}$; 
\item[$\bullet$] Compute the singular value decomposition of $E_{i,j}^{(n)} = \sum_{q=0}^{n-1} (E_i E_j)^q$, and obtain $E_{i,j}^{(n)} = U_{i,j}\Lambda_{i,j} V_{i,j}$, where $U_{i,j}$ and $V_{i,j}$ are unitary matrices, and $\Lambda$ is a diagonal matrix; 
\item[$\bullet$] Suppose $\lambda_{i,j;{\rm max}}^{(n)}$ is the largest singular value of $E_{i,j}^{(n)}$, construct the equation 
\begin{eqnarray}
V_{i,j;1,\bullet} \vec{k}_{i,j} = \lambda_{i,j;{\rm max}}^{(n)-1} \left(U_{i,j}^{-1} \vec{k}_{i,j}^{(n)}\right)_1
\label{eq:double}
\end{eqnarray}
for each $(i,j)$, where $\vec{k}_{i,j} = \vec{k}_{i} + E_i \vec{k}_{j}$ is the non-unital part of $M_i M_j$. We assume that the first singular value is the largest one, i.e.~$\Lambda_{1,1} = \lambda_{i,j;{\rm max}}^{(n)}$, then $V_{i,j;1,\bullet}$ is the first row of $V_{i,j}$, and $\left(\bullet\right)_1$ denotes the first element of the vector; 
\item[$\bullet$] Solve SOE~(\ref{eq:double}) to obtain the non-unital part $\vec{k}_{i}$ of each gate. 
\end{itemize}
Given $N$ gates in the gate set, we can construct at most $3N-3$ linearly independent equations, where $3$ is due to the gauge freedom in GST, similar to the unital part. See Appendix~\ref{app:gauge}. We numerically find that $21$ double maps in the form $M_i M_j$ can generate $3N-3$ linearly independent equations. Here, $i<j$, and $M_i$ and $M_j$ are gates in Table~\ref{table:gate_set}. 

\section{Numerical Simulation}
\label{sec:num}

In this section, we demonstrate our protocol with the numerical simulation. We use the gate set given in Table~\ref{table:gate_set}. For each ideal gate $M_j^{\rm i}$, where $j=1,2,\ldots,7$, we randomly generate the corresponding noisy gate $M_j$ following the approach in Appendix~\ref{app:error}. Then, we use our protocol to reconstruct the noisy gates for the gate set. 

To estimate the trace of a map $M$, we solve SOE~(\ref{eq:trace}) for a monotonically increasing sequence $n = m\lfloor 2^k/m \rfloor + 1$, where $k = 0,1,\ldots,\lfloor \log_2(0.4/p) \rfloor$, and $m$ is the period of $M^{\rm i}$, i.e.~the smallest positive integer such that ${M^{\rm i}}^m = \openone$. When $k = 1$, we have only one solution of equations. When $k > 1$, there are multiple solutions, and we always choose the one that is closest to the solution in the previous step. In this way, we can eventually determine the solution of $k = \lfloor \log_2(0.4/p) \rfloor$, which is used to compute the trace of the map. In our protocol, each $t_{l} = \Tr(\hat{M}_{l})-1$ in the equations is measured using GST. In our simulation, we take $\Tr(\hat{M}_{l}) = \Tr(M^l) + \zeta$, where $\zeta$ is a random number generated according to the normal distribution with zero mean and the standard deviation $\sigma = 0.01$ that represents the sampling noise. This standard deviation means that each diagonal element of $\hat{M}_{l}$ is measured with the accuracy $\sim 0.01/4 = 0.0025$ in GST. 

To obtain the unital part of maps, we use a hundred quadruple maps listed in Table~\ref{table:quadruple} to construct a hundred equations according to Eq.~(\ref{eq:quadruple}), in which $\Tr(M_{i,j,k,l})$ is measured using the trace measurement. SOE of the unital part has the rank of $55$ and $63$ unknown variables. We determine the solution using the Moore-Penrose inverse~\cite{ben2003generalized}: We take $\mathbf{x} = A^+\mathbf{b}$ as the solution of the equation $A\mathbf{x} = \mathbf{b}$, where $A^+$ is the Moore-Penrose inverse of $A$. 

\begin{figure}[tbp]
\centering
\includegraphics[width=1\linewidth]{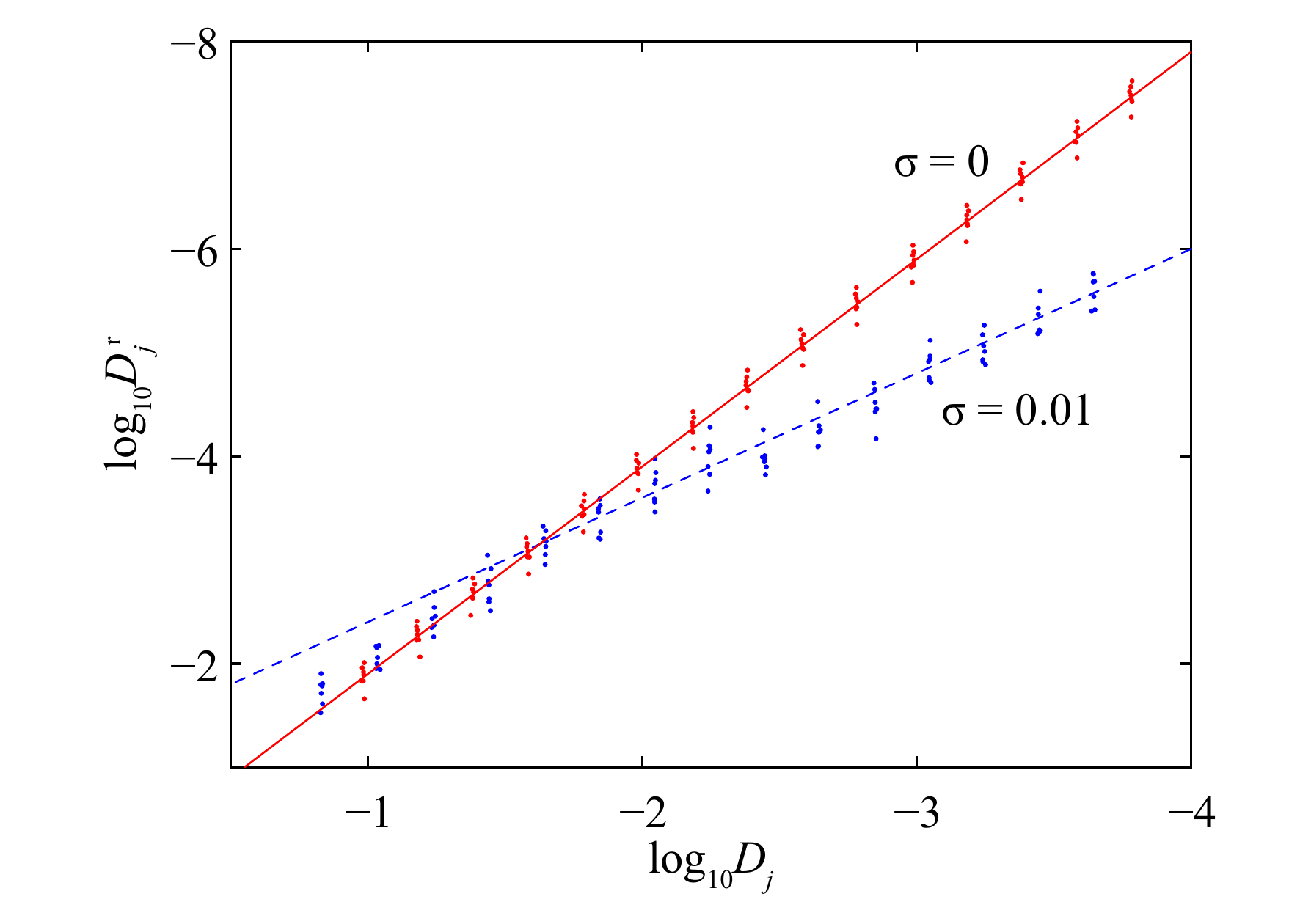}
\caption{
Distances of randomly generated noisy gates sets. Thirty gate sets are generated. Fifteen of them are reconstructed in the numerical simulation taking the sampling noise $\sigma = 0.01$ (blue dots), and the other fifteen gate sets are reconstructed taking $\sigma = 0$ (red dots). Error rates are in the range $\sim 10^{-3} - 10^{-6}$. Straight lines are $\log_{10}D^{\rm r}_j = 1.2\log_{10}D_j - 1.2$ (blue) and $\log_{10}D^{\rm r}_j = 2\log_{10}D_j + 0.1$ (red), respectively. The slop grater than one means that the relative error decreases with the distance. 
}
\label{fig:distance}
\end{figure}

To use the result of the unital part in the reconstruction of the non-unital part, we need to find a proper similarity transformation. The unital part obtained using our protocol, which is denoted by $\hat{E}$, has an unknown similarity transformation from the actual unital part, i.e.~$\hat{E} = BE{B}^{-1}$ (neglecting the sampling noise and higher-order effects in the perturbation). The matrix $B$ depends on how we choose the solution of Eq.~(\ref{eq:quadruple}). In the reconstruction of the non-unital part, the non-unital part of maps $(M_iM_j)^n$ is measured using GST, and there is an unknown transformation from the actual non-unital part, i.e.~$\hat{\vec{k}}' = B'\vec{k}+\vec{a}'-B'E{B'}^{-1}\vec{a}'$ is the result of GST. Here, the matrix $B'$ depends on details of GST, including the state preparation and measurement error. Therefore, two matrices $B$ and $B'$ are different in general. We need to find a proper similarity transformation relates the result of SOE~(\ref{eq:quadruple}) to the result of GST. Under the assumption that transformations from the actual map is close to identity, we can find the proper similarity transformation by solving equations. See Appendix~\ref{app:transformation1} for details. \\
\vspace{-2pt}
In the reconstruction of the non-unital part, we first measure $21$ maps $(M_iM_j)^n$ ($i<j$) using GST, where $n = \lfloor 1/p_{i,j} \rfloor$, where $p_{i,j}$ is the error rate of $M_iM_j$. The result is also used to determine the similarity transformation. In the numerical simulation, we take the result of the map $(M_iM_j)^n$ as $T (M_iM_j)^n T^{-1} + \zeta$, where $T$ is a randomly generated matrix representing unknown transformation from the actual map, and $\zeta$ is a matrix represents the sampling noise. $T$ is generated using the same approach for generating the noise in an actual map, and we take the error rate $p = 0.1$. See Appendix~\ref{app:error}. Each element of $\zeta$ is generated according to the normal distribution with the zero mean and the standard deviation $\sigma = 0.01$. Using the largest singular value of each double map, we have $21$ equations. The system of equations have $21$ unknown variables, corresponding to the non-unital part of the seven gates. However, three singular values of the system of equations (\ref{eq:double}) are small. To obtain a stable solution, we apply the truncation on singular values, i.e.~replace the three small singular values with zero, and then determine the solution using the Moore-Penrose inverse.\\
\vspace{-2pt}
To demonstrate that we can reconstruct high-fidelity gates with our protocol, we compare the reconstructed maps with actual maps. We use $M_j^{\rm r}$ to denote the reconstructed map. Because of the gauge problem, maps $M_j^{\rm r}$ and $M_j$ cannot be directly compared. Even our protocol is implemented ideally, the reconstruction is still up to an unknown similarity transformation, i.e.~$TM_j^{\rm r}T^{-1} = M_j$. The matrix $T$ cannot be determined in GST because of the state preparation and measurement errors~\cite{greenbaum2015introduction}. It is the same in our protocol. Therefore, the reconstruction is successful if there is a matrix $T$ such that $TM_j^{\rm r}T^{-1} - M_j$ is small for all $j$. We can find the matrix $T$ as shown in Appendix~\ref{app:transformation2}. The result of $D^{\rm r}_j = \norm{TM_j^{\rm r}T^{-1} - M_j}_2$ for noisy gate sets with different error rates are plotted in Fig.~\ref{fig:distance}. We can find that the relative error of the reconstruction $D^{\rm r}_j / D_j$ decreases with $D_j$, where the distance $D_j = \norm{M_j - M_j^{\rm i}}_2$ measures the error in the gate. Comparing results of the sampling noise $\sigma = 0.01$ to the case without sampling noise, we can find that the sampling noise reduces the reconstruction accuracy when $D_j$ is smaller than $0.01$. \\
\vspace{-2pt}
In our numerical simulation, we have use the prior knowledge of the gate error rate, such that we can choose the proper number of gate repetitions. In the practical implementation, we can choose the proper repetition number by measuring gate sequences for a set of repetition numbers, e.g.~increasing the repetition number exponentially such as in the trace measurement. We note that the performance is not sensitive to the repetition number as shown in Figs.~\ref{fig:trace_variance}(a)~and~\ref{fig:singular_value}. 
\section{conclusion}
\label{sec:conclusion}

In this paper we propose a protocol to reconstruct unknown quantum gates with high fidelity. This method reduces the impact of sampling noise by amplifying the error in deterministic gate sequences. Compared with analyzing data of gate sequences using the maximum likelihood estimation~\cite{blume2017demonstration}, our approach is based on solving linear equations rather than optimization algorithm. We can improve the accuracy of reconstruction by using the maximum likelihood estimation method and taking the result of our perturbative approach as the initial estimate of the error model. We demonstrate our protocol in numerical simulation and find that the relative error of reconstruction decreases with the gate error. Because our approach includes increasing the gate repetition number exponentially to approximately one over the error rate in the unital part reconstruction, the number of measurements needed in our protocol scales logarithmically with the error rate. However, because we need to amplify the error in sufficiently long gate sequences, the number of gates scales linearly. As long as the time cost of implementing gate sequences is practical, our protocol provides a way to choose proper gate sequences and efficiently reconstruct high fidelity quantum gates. 

Our code used for generating numerical data in this paper can be found at \href{https://github.com/yangruyu96/fast_GST}{code}.
\begin{acknowledgments}
This work is supported by National Natural Science Foundation of China (Grant No. 11875050) and NSAF (Grant No. U1930403).
\end{acknowledgments}

\normalem

\appendix

\section{Random error generation}
\label{app:error}

Given the ideal map $\mathcal{M}^{\rm i}$ and the error rate $p$, we generate the map with error as follows. The Lindblad equation for a single qubit can be written as $\frac{d\rho}{dt} = \mathcal{L}(\rho)$, and 
\begin{eqnarray}
\mathcal{L}(\rho) &=& -i[H,\rho] \notag \\
&&+ \sum_{a,b=1}^3 h_{a,b} \left[\sigma_a\rho\sigma_b - \frac{1}{2}(\sigma_b\sigma_a\rho+\rho\sigma_b\sigma_a)\right]
\end{eqnarray}
Here, $H$ is the Hamiltonian, $h$ is a positive semidefinite matrix, and $\sigma_a$ are Pauli operators. To generate the error, we first randomly generate $H$ and $h$. The map with error is $\mathcal{M} = e^{\mathcal{L}t} \mathcal{M}^{\rm i}$, where $e^{\mathcal{L}t}$ represents the noise. By choosing the evolution time $t$, we can obtain the map with the desired error rate $p$. We use the same approach to generate the matrix $T$ of GST, by taking $T$ as the Pauli transfer matrix of $e^{\mathcal{L}t}$. 

\section{Gauge freedom}
\label{app:gauge}

According to the GST formalism, we can only determine the map in the tomography experiment up to a similarity transformation, i.e.~two sets of maps $\{M_i\}$ and $\{T M_i T^{-1}\}$ are indistinguishable. All maps $\{M_i\}$ and $\{T M_i T^{-1}\}$ are in the form of Eq.~(\ref{eq:M}), i.e.~the first row is $(1,0,0,0)$, which sets four constraint conditions on $T$. Therefore, we can express $T$ as 
\begin{eqnarray}
T = \left[
\begin{array}{cc}
1 & \vec{0}^{\rm \, T}\\
\vec{a} & B
\end{array}
\right].
\label{eq:T}
\end{eqnarray}
We take the first element as one, because similarity transformations given by $T$ and $\alpha T$ are the same, where $\alpha$ is a non-zero scalar factor. The inverse matrix is 
\begin{eqnarray}
T^{-1} = \left[
\begin{array}{cc}
1 & \vec{0}^{\rm \, T}\\
B^{-1}\vec{a} & B^{-1}
\end{array}
\right].
\end{eqnarray}
After the similarity transformation, we have 
\begin{eqnarray}
TMT^{-1} = \left[
\begin{array}{cc}
1 & \vec{0}^{\rm \, T}\\
B\vec{k}+\vec{a}-BEB^{-1}\vec{a} & BEB^{-1}
\end{array}
\right].
\end{eqnarray}

We can find that the similarity transformation of $T$ causes a similarity transformation on the unital part, i.e.~$E\rightarrow BEB^{-1}$. The matrix $B$ is $3\times 3$ and has $9$ elements. The similarity transformation is invariant when the matrix is scaled by a non-zero scalar factor. Therefore, the similarity transformation of the unital part has $8$ degrees of freedom, e.g.~$8$ parameters cannot be determined in the reconstruction of the unital part. 

To be specific, in our perturbative approach, we assume that $\delta M$ is small, which implies that $T$ is close to identity. Therefore, $\vec{a}$ and $\delta B \equiv B - \openone$ are small. The inverse matrix of $B$ is approximately $B^{-1} \simeq \openone - \delta B$. If we neglect high order terms, the error after the similarity transformation is 
\begin{eqnarray}
BEB^\dag - E^{\rm i} \simeq \delta E + \delta BE^{\rm i} - E^{\rm i}\delta B.
\end{eqnarray}
We can find that if $\{ \delta E_j \}$ is the solution of Eq.~(\ref{eq:quadruple}), $\{ \delta E_j + \delta BE_j^{\rm i} - E_j^{\rm i}\delta B \}$ is also a solution. If we replace $\delta B$ with $\delta B + \alpha \openone$, where $\alpha$ is a scalar factor, the solution does not change. Therefore, there are $8$ non-trivial degrees of freedom. 

The non-unital part after the similarity transformation is approximately $\vec{k} + \vec{a} - E^{\rm i} \vec{a}$. Here we have used that $B \approx \openone$ and $E \approx E^{\rm i}$. $E^{\rm i}$ is a unitary matrix, and one of its eigenvalues is one, which corresponds to the largest singular value in the non-unital part reconstruction. We only use the largest singular value in the non-unital part reconstruction, i.e.~the non-unital part component that contributes to the reconstruction is $P(\vec{k} + \vec{a} - E^{\rm i} \vec{a})$, where $P$ is the projection operator onto the eigenvector (with the eigenvalue one) of $E^{\rm i}$. We can find that $P(\vec{k} + \vec{a} - E^{\rm i} \vec{a}) = P\vec{k}$, i.e.~if $\vec{k}$ is the solution to the equation of the non-unital part, $\vec{k} + \vec{a} - E^{\rm i} \vec{a}$ is also a solution. Because $\vec{a}$ has three elements, $3$ parameters cannot be determined in the reconstruction of the non-unital part. 

We remark that in the discussion of the non-unital part, we have taken the approximations $B \approx \openone$ and $E \approx E^{\rm i}$. Because of the finite error in $B$ and $E$, we can find more than $3N-3$ linearly independent equations. However, the system of equations for the non-unital part has up to $3N-3$ singular values that are reasonably large. 

\subsection{Compute the transformation - Non-unital part}
\label{app:transformation1}

Let $\hat{E}$ and $\hat{E}'$ be unital parts obtained by solving SOE~(\ref{eq:quadruple}) and GST, respectively. There are similarity transformations relate them to the actual unital part $E$, i.e.~$\hat{E} = BE{B}^{-1}$ and $\hat{E}' = B'E{B'}^{-1}$. Here, we have assumed that $\hat{E}$ and $\hat{E}'$ are obtained without the sampling noise. We want to find $B'' = B{B'}^{-1}$ such that we can compute $\hat{E}'' = B''\hat{E}'{B''}^{-1}$. Ideally, we have $\hat{E}'' = \hat{E}$. 

Under the condition that the error in gates is small, maps obtained by solving SOE~(\ref{eq:quadruple}) and GST are both close to the ideal map. Therefore, matrices $B$, $B'$ and $B''$ must be close to the identity matrix. We assume that $B'' = \openone + \delta B$ and $\delta B$ is small. 

Let $\hat{E}_{i}$ be the unital part of $M_i$ obtained by solving SOE~(\ref{eq:quadruple}). We compute $Y_{i,j} = (\hat{E}_{i}\hat{E}_{j})^n$. Let $X_{i,j}$ be the unital part of $(M_iM_j)^n$ measured using GST. Then, we have equations 
\begin{eqnarray}
\delta B X_{i,j} - X_{i,j} \delta B = Y_{i,j}.
\label{eq:deltaB1}
\end{eqnarray}
Here, we have neglected high-order terms of $\delta B$. We have $21$ double maps, therefore, $21\times 9 = 189$ equations. SOE~(\ref{eq:deltaB1}) has $9$ unknown variables, but the rank is $8$. The variable that cannot be determined corresponds to scaling the similarity transformation matrix by a non-zero scalar factor, which is trivial. We solve SOE~(\ref{eq:deltaB1}) using the Moore-Penrose inverse. 

With the matrix $\delta B$, we compute $B'' = \openone + \delta B$ and $\hat{\vec{k}} = B''\hat{\vec{k}}'$. Then, $\hat{\vec{k}}$ is used as $\vec{k}$ ($\vec{k}_{i,j}^{(n)}$) in the reconstruction of the non-unital part. 

\subsection{Compute the transformation - Benchmarking}
\label{app:transformation2}

To compute the similarity transformation that relates $M_j^{\rm r}$ to $M_j$, we assume $T M_j^{\rm r} T^{-1} = M_j$, and $T$ is in the form given by Eq.~(\ref{eq:T}). We assume $T$ is close to identity, i.e.~$\vec{a}$ and $\delta B = B - \openone$ are small. 

To compute $B$, we solve the equations 
\begin{eqnarray}
\delta B E_j^{\rm r} - E_j^{\rm r} \delta B = E_j.
\label{eq:deltaB2}
\end{eqnarray}
Here, $E_j^{\rm r}$ is the unital part of the reconstructed gate $M_j^{\rm r}$, and $j = 1,2,\ldots,7$. Here, we have neglected high-order terms of $\delta B$. We have $7$ maps, therefore, $7\times 9 = 63$ equations. As the same as in the case of SOE~(\ref{eq:deltaB1}), there are $9$ unknown variables, but the rank is $8$. We solve SOE~(\ref{eq:deltaB2}) using the Moore-Penrose inverse. 

Given $\delta B$ and $B = \openone + \delta B$, we have equations 
\begin{eqnarray}
B\vec{k}_j^{\rm r}+\vec{a}-BE_j^{\rm r}B^{-1}\vec{a} = \vec{k}_j,
\label{eq:a}
\end{eqnarray}
where $\vec{k}_j^{\rm r}$ is the non-unital part of $M_j^{\rm r}$. We have $7$ maps, therefore, $7\times 3 = 21$ equations. There are $3$ unknown variables. We solve SOE~(\ref{eq:a}) using the Moore-Penrose inverse. \\
\vspace{-2pt}
In the computation of $T$, we assume that actual maps $M_j$ are known, which is only used for benchmarking the result in the numerical simulation and not needed in the implementation of our protocol. 
\bibliography{reference}
\end{document}